\documentclass[conference,a4paper]{IEEEtran}
\IEEEoverridecommandlockouts

\makeatletter
\def\markboth#1#2{\def\leftmark{\@IEEEcompsoconly{\sffamily}\MakeUppercase{\protect#1}}%
\def\rightmark{\@IEEEcompsoconly{\sffamily}\MakeUppercase{\protect#2}}}
\makeatother

\newcommand\blfootnote[1]{%
  \begingroup
  \renewcommand\thefootnote{}\footnote{#1}%
  \addtocounter{footnote}{-1}%
  \endgroup
}

\usepackage[english]{babel}
\usepackage{xcolor}
\usepackage{lipsum}
\usepackage{caption}
\usepackage{cite}
\usepackage[pdftex]{graphicx}
\usepackage{subcaption}
\usepackage{amsmath}
\usepackage{booktabs}
\usepackage{colortbl}
\usepackage{amsfonts}
\usepackage{amssymb}
\usepackage{amsthm}
\usepackage{array}
\usepackage{verbatim}
\usepackage{listings}
\usepackage{algorithm}
\usepackage{algpseudocode}
\usepackage{url}
\usepackage{enumerate}
\usepackage{multirow}

\usepackage{epsfig}
\usepackage{epstopdf}
\usepackage{multicol}
\usepackage[font=footnotesize]{caption}

\usepackage{hyperref}

\newcommand{\bi}{\begin{itemize}}
\newcommand{\ei}{\end{itemize}}
\newcommand{\be}{\begin{equation}}
\newcommand{\ee}{\end{equation}}

\def\beq{\begin{equation}}
\def\eeq{\end{equation}}
\def\beqa{\begin{eqnarray}}
\def\eeqa{\end{eqnarray}}
\def\beqan{\begin{eqnarray*}}
\def\eeqan{\end{eqnarray*}}

\title{Understanding Noise and Interference Regimes\\ in 5G Millimeter-Wave Cellular Networks}
\author{{{\bf Mattia Rebato}$^\dagger$, {\bf Marco Mezzavilla}$^*$, {\bf Sundeep Rangan}$^*$, {\textbf{Federico Boccardi}$^\diamond$}, {\bf Michele Zorzi}$^\dagger$ }\\
$^*$ NYU WIRELESS, Brooklyn, NY, USA \qquad\qquad 
$^\dagger$ University of Padova, Italy\\
emails: \small{$\{$\texttt{rebatoma}, \texttt{zorzi}$\}$\texttt{@dei.unipd.it}, $\{$\texttt{mezzavilla}, \texttt{srangan}$\}$\texttt{@nyu.edu}, \texttt{federico.boccardi@ieee.org}
}}

\begin{document}
\maketitle

\begin{abstract}
With the severe spectrum shortage in conventional cellular bands, millimeter-wave (mmWave) frequencies have been attracting growing attention
for next-generation micro- and pico-cellular wireless networks.
A fundamental and open question is whether mmWave cellular networks 
are likely to be  noise- or interference-limited.
Identifying in which regime a network is operating is critical 
for the design of MAC and physical-layer procedures and to provide insights on how transmissions across cells should be coordinated to cope with interference.
This work uses the latest measurement-based statistical channel models to accurately assess the Interference-to-Noise Ratio (INR) in a wide range of deployment scenarios.
In addition to cell density, we also study antenna array size and antenna patterns, whose effects are critical in the mmWave regime.  
The channel models also account for blockage, line-of-sight and non-line-of-sight regimes as well as local scattering, that significantly affect the level of spatial isolation.\blfootnote{$^\diamond$ F. Boccardi's work was carried out in his personal capacity and the views expressed here are his own and do not reflect those of his employer (Ofcom).}
\end{abstract}
\smallskip
\begin{IEEEkeywords}
5G, millimeter wave communication, cellular systems, 
interference regime, noise regime
\end{IEEEkeywords}

\section{Introduction}
\label{introduction}
The millimeter-wave (mmWave) spectrum, roughly defined as the frequencies between 10 and 300~GHz,  is a new and promising frontier for cellular wireless communications~\cite{ted_book,ted4}.
With the rapidly growing demand for cellular data, conventional frequencies below 3~GHz are now highly congested. For example, in the most recent FCC auction,
65~MHz of AWS-3 spectrum were sold for a record breaking \$45 billion, which shows the severe spectrum crunch encountered when trying to expand wireless networks today. In contrast, the mmWave bands offer vast and largely untapped spectrum, up to 200 times all current cellular allocations by some estimates.
Due to this enormous potential, mmWave networks have been widely cited as one of the most promising technologies for Beyond 4G and 5G cellular evolution.

A fundamental and outstanding question for the design of these
networks is to understand the effects of interference and, more
specifically, under which circumstances mmWave cellular 
networks are likely to be limited by interference or by thermal noise.
Identifying in which regimes networks operate is central to system design:
for example, while interference limited networks can benefit from 
advanced techniques such as
inter-cellular interference coordination, coordinated beamforming
and dynamic orthogonalization, these techniques have little value
in networks where thermal noise, rather than interference, is dominant.  

While for traditional (in particular macrocell-based) cellular deployments the relative power 
of interference to thermal noise
is a function of the distance between cells and of the
transmit power spectral density, the results in this paper will demonstrate that in mmWave systems
the relative strength of
interference depends on many more factors.  Most importantly,
mmWave systems rely on highly directional transmissions to overcome
the high isotropic path loss.  Directional transmissions
tend to isolate users, thereby reducing the interference.  
However, the degree of isolation depends strongly on the size of the antenna arrays, the antenna pattern, 
and the level of local scattering and spatial multipath.
In addition, mmWave signals can be blocked by many common materials,
eliminating long distance links.  This potentially improves the isolation
but may also lead to coverage holes.  There are also significant differences
in path loss for mobiles in Line-of-Sight (LoS) and Non-Line-of-Sight
(NLoS) locations.  

The broad purpose of this paper is to leverage detailed measurement-based
statistical channel models to provide an accurate assessment of 
the Interference-to-Noise Ratio (INR), and of its relation to various
key deployment parameters including base station density,
transmit power, bandwidth and antenna pattern.  Our analysis uses
the latest channel models for 28 and 73~GHz based on extensive New York City
measurements~\cite{ted1,ted2,ted3,ted4}.

The rest of the paper is organized as follows. In Section \ref{related_work} we present the prior art related to interference and noise evaluations.
In Section \ref{system_model} we describe the scenarios simulated.
A preliminary numerical evaluation, along with some important remarks, is reported in Section \ref{numerical_evaluation}. Finally, we conclude the paper and describe some future research steps in Section \ref{conclusion}.

\section{Related work}
\label{related_work}

In~\cite{management1} and~\cite{management2}, the authors outline the challenges of interference management in 5G cellular networks, overviewing techniques such as Coordinated MultiPoint (CoMP) and other advanced interference management techniques.

In~\cite{renzo}, a new mathematical framework for the analysis of mmWave cellular networks is presented.
The paper introduces a multi-ball approximation that lies in replacing the LoS or NLoS probability of typical User Equipment (UE) with an approximate function, which still depends on the Base Station (BS) to UE distance $d$ but is piece-wise constant as a function of $d$. A noise-limited approximation is found to be quite accurate for a small density of BSs. However, when the density of BSs increases, thus decreasing the average cell radius, the approximation no longer holds.

Noise-limited and interference-limited regimes are studied for ad hoc mmWave networks at $60$~GHz operating under slotted ALOHA and Time Division Multiple Access (TDMA) protocols in~\cite{fischione} and~\cite{fischione2}.
These works consider networks at 60~GHz, motivated by the fact that a lot of research has been done in the last years to support the development of WiGig~\cite{wigig}. 
However, we note that propagation at 60~GHz is heavily affected by the peak of oxygen absorption, and for this reason different Access Points (APs) will be more isolated than at 28 or 73~GHz.
Different scenarios are studied and implemented from both analytical and simulation standpoints.
As a result, \cite{fischione} and~\cite{fischione2} suggest the use of a hybrid MAC protocol that works in two distinct phases; a distributed contention-based resource allocation, which is more suitable for the noise-limited regime, followed by a centralized contention-free resource allocation, which is more suitable for the interference-limited regime.

In~\cite{heath}, an analytical framework is proposed to evaluate the instantaneous INR distribution of an outdoor mmWave ad hoc network working at 60~GHz.
The authors consider a narrowband channel model with transmitter locations forming a Poisson Point Process (PPP), and capture mmWave features by considering directional beamforming and LoS/NLoS configurations. It is shown that, in a very dense network (e.g., 1000~sources/km$^2$), the interference power is nearly always higher than the noise power.
This motivates novel ad hoc mmWave architectures to deal with interference in order to realize networks that can achieve gigabit speeds.

Some general results related to interference and noise regimes for mmWave cellular networks have been presented in~\cite{sundeep_challenges} and~\cite{heath2}, which motivated us to run a more detailed campaign of simulations at varying operating configurations.  The work in 
\cite{felipe} used simple approximations of the channel 
propagation to identify scaling laws for the bandwidth, number of antennas
and transmit power under which the network would be in an interference or
noise-limited regime.

Reference~\cite{gupta15} studies the feasibility of spectrum pooling in mmWave cellular networks under ideal conditions at both 28 and 73~GHz, and \cite{boccardi16} studies the impact of coordination between different networks, under the assumption of ideal beamforming.
In both~\cite{gupta15} and~\cite{boccardi16}, beamforming is modeled as an ON/OFF beam.

Our INR evaluation introduces some key additional contributions: (i) a detailed \emph{lobe}-shaped antenna pattern to precisely capture the mmWave beamforming gains\footnote{A precise description of the model used in this paper can be found in~\cite{rebato16}.}; (ii) an updated channel model at both 28 and 73~GHz\footnote{We use the NYU channel models for 28~GHz and 73~GHz frequencies~\cite{mustafa} based on measurement campaigns carried out in a real dense urban environment, as reported in~\cite{ted1,ted2,ted3,ted4}.}, which, together with a more realistic beamforming model, provides a very accurate characterization of the useful and interfering received signal; and (iii) a comparison of a blind allocation vs. a centralized upper bound vs. an interference-less case.
We believe that these additional contributions,  with respect to the previous work, are an important step forward towards a more realistic understanding of the role of interference in 5G mmWave system design.

\section{System model}
\label{system_model}

We consider a mobile network where BSs and UEs are deployed following a PPP with density $\lambda_{\text{BS}}$ and $\lambda_{\text{UE}}$, respectively.
Moreover, the use of this unplanned deployment where the BS positions are not optimized is suitable to model the case where APs are deployed by users in a similar way to WiFi today.
Our simulations follow a Monte Carlo approach, in which many independent experiments\footnote{More precisely, experiments are independent because the deployment of the devices in the area is randomly generated at each iteration.} are repeated to empirically derive statistical quantities of interest. Without loss of generality, we evaluate the performance of a typical receiver located in the origin of the area considered, whose statistics are estimated based on 50000 repetitions of this procedure.

Thanks to a detailed channel and antenna characterization, we can compute the Signal-to-Interference-plus-Noise-Ratio (SINR) between transmitter $i$ and receiver $j$ as:
\be
\text{SINR}_{ij} = \frac{\frac{P_{Tx}}{PL_{ij}}G_{ij}}{\sum_{k \neq i} \frac{P_{Tx}}{PL_{kj}}G_{kj} + BW \times N_0},
\label{equation_sinr}
\ee
where $k$ represents each interfering link, $BW$ is the total bandwidth, $N_0$ is the thermal noise, $P_{Tx}$ is the transmitted power, $G$ is the beamforming gain, and $PL$ is the pathloss between the receiver UE and the associated BS.

The pathloss is modeled with three states, as reported in~\cite{mustafa}: LoS, NLoS and outage as a function of the distance $d$ between transmitter and receiver. 
In the simulations, UEs are associated to the BS that provides the smallest pathloss. 
The channel is modeled as reported in~\cite{mustafa}, which represents a dense urban environment. 
A precise \emph{lobe}-shaped beamforming gain $G$ is computed by multiplying the Multiple-Input and Multiple-Output (MIMO) beamforming vectors\footnote{The beamforming vectors are computed as reported in \cite{antenna_book}. This antenna gain model is detailed and complete with main and side-lobes.} by the channel matrix.
The beamforming gain from transmitter $i$ to receiver $j$ is given by:
\be
G_{ij} = |\textbf{w}^*_{Rx_{ij}} \textbf{H}_{ij} \textbf{w}_{Tx_{ij}} |^2,
\label{bf_comp}
\ee
where $\textbf{w}_{Tx_{ij}} \in \mathbb{C}^{n_{Tx}}$ is the beamforming vector of transmitter $i$ when transmitting to receiver $j$, $\textbf{w}_{Rx_{ij}} \in \mathbb{C}^{n_{Rx}}$ is the beamforming vector of receiver $j$ when receiving from transmitter $i$, and $\textbf{H}$ is the channel matrix. 
Both vectors are complex, with length equal to the number of antenna elements in the array.
The use of beamforming is essential in mmWave communications, as the gain obtained from directional beam steering is a critical factor to achieve a sufficient link margin.
We assume to have the possibility of steering in any direction, i.e., we can generate a beamforming vector for any possible angle between 0 and 360 degrees.
With this transmission, the two beams (i.e., the one of the UE and the one of the BS) are always perfectly
aligned. 
From a realistic point of view, the set of beampatterns is discrete, and performing the steering in an arbitrary direction may be too costly or even impossible. 

\section{Simulation results}
\label{numerical_evaluation}
\begin{figure}[t!]
\centering
\includegraphics[width=0.77\columnwidth]{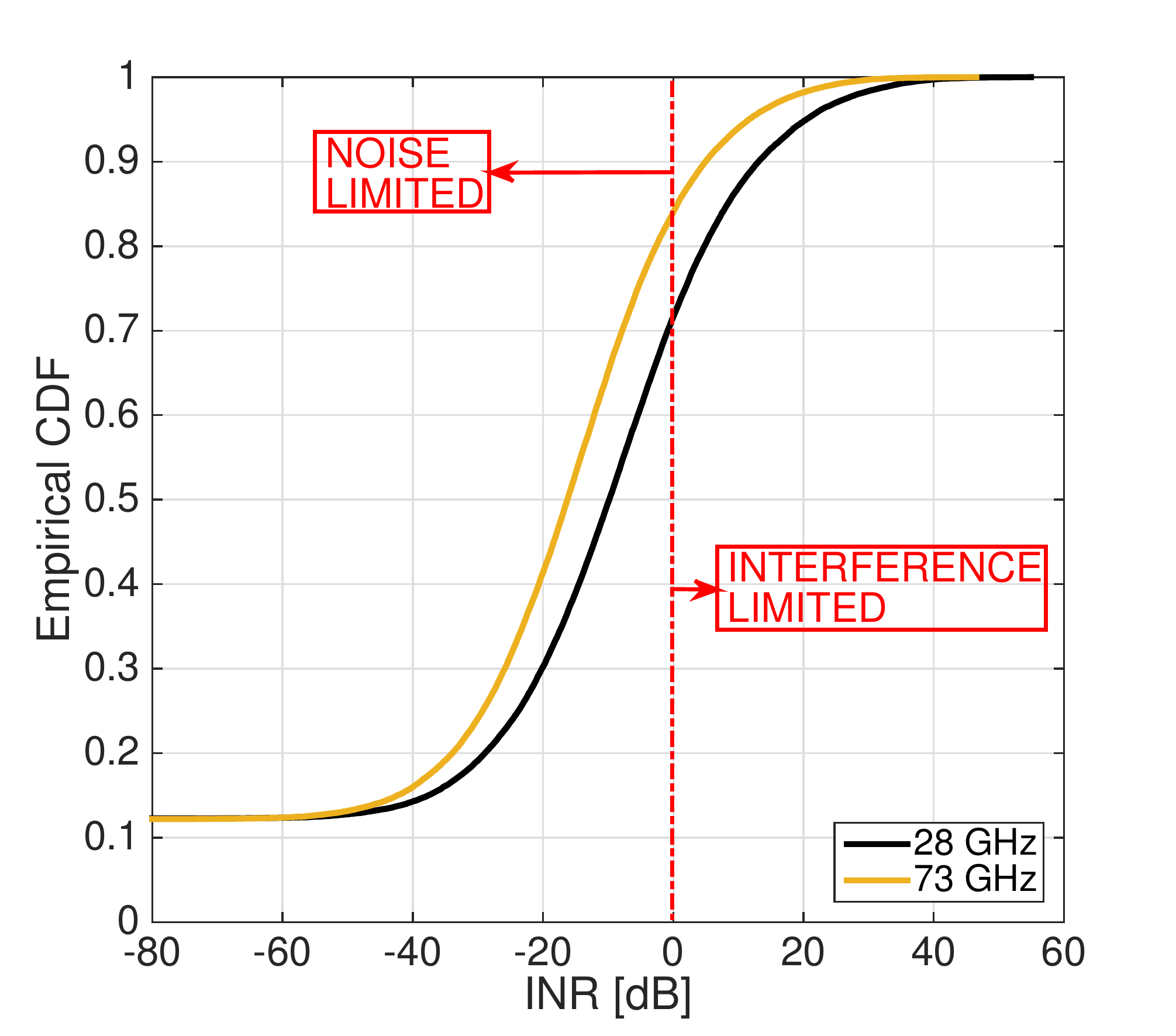}     
\caption{Empirical CDF of the INR for $\lambda_{\text{UE}}$ = 300~UEs/km$^2$ and $\lambda_{\text{BS}}$ = 30~BSs/km$^2$.}   
\label{300}
\end{figure}
\begin{figure}[t!]
\centering
\includegraphics[width=0.77\columnwidth]{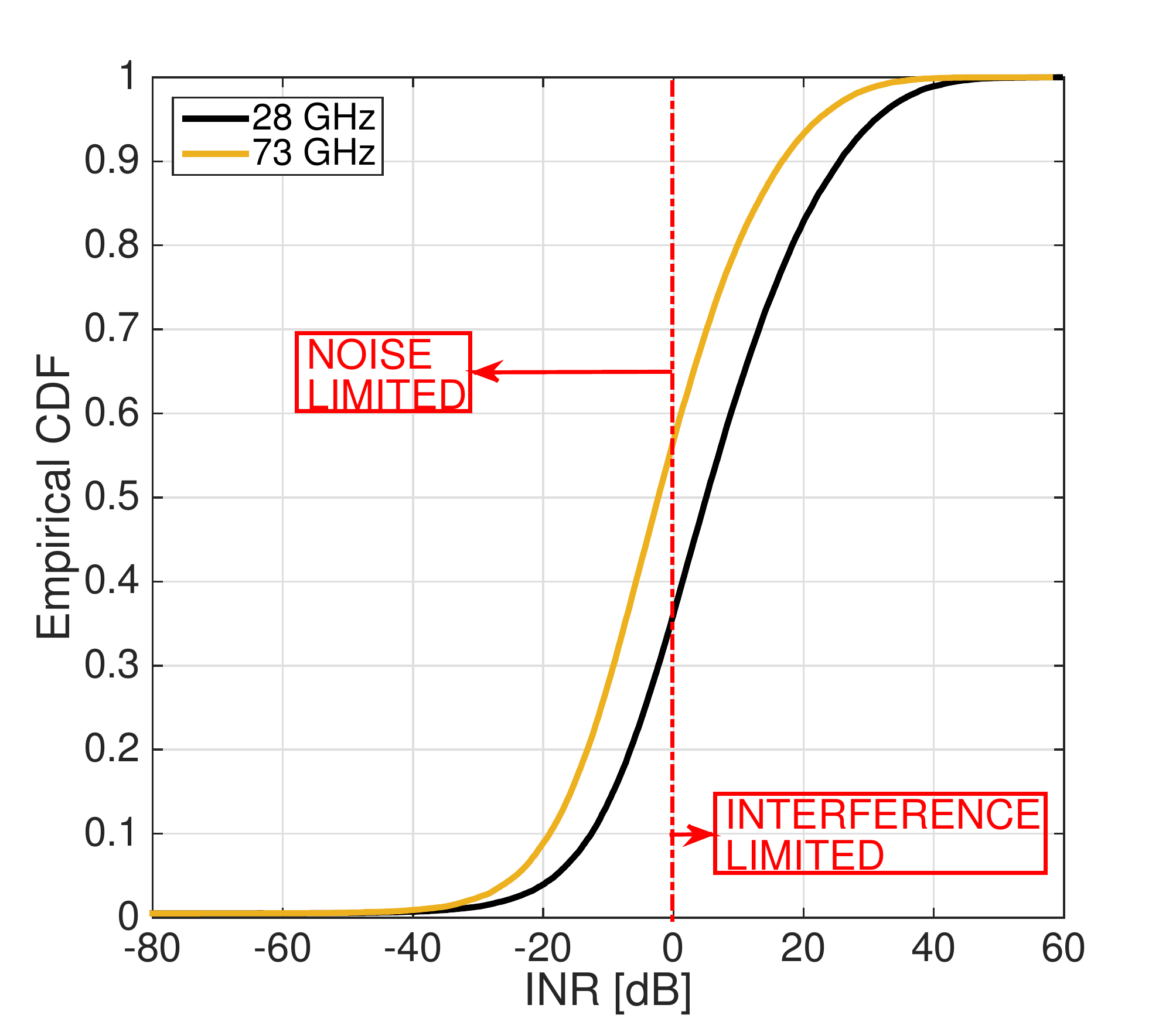}     
\caption{Empirical CDF of the INR for $\lambda_{\text{UE}}$ = 600~UEs/km$^2$ and $\lambda_{\text{BS}}$ = 60~BSs/km$^2$.}   
\label{600}
\end{figure}
\begin{figure}[t!]
\centering
\includegraphics[width=0.77\columnwidth]{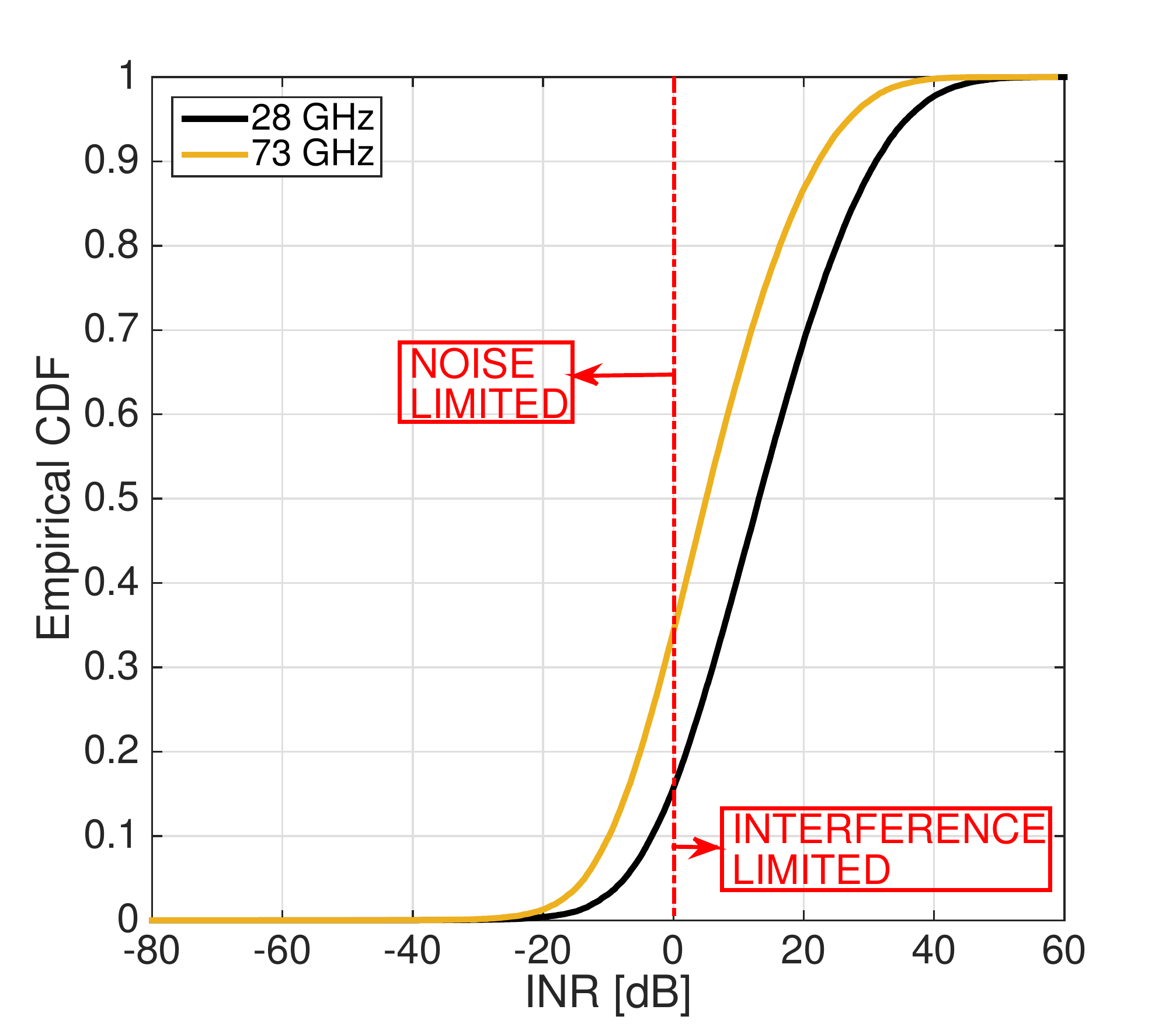}     
\caption{Empirical CDF of the INR for $\lambda_{\text{UE}}$ = 900~UEs/km$^2$ and $\lambda_{\text{BS}}$ = 90~BSs/km$^2$.}   
\label{900}
\end{figure}
\begin{figure}[t!]
\centering
\includegraphics[width=0.77\columnwidth]{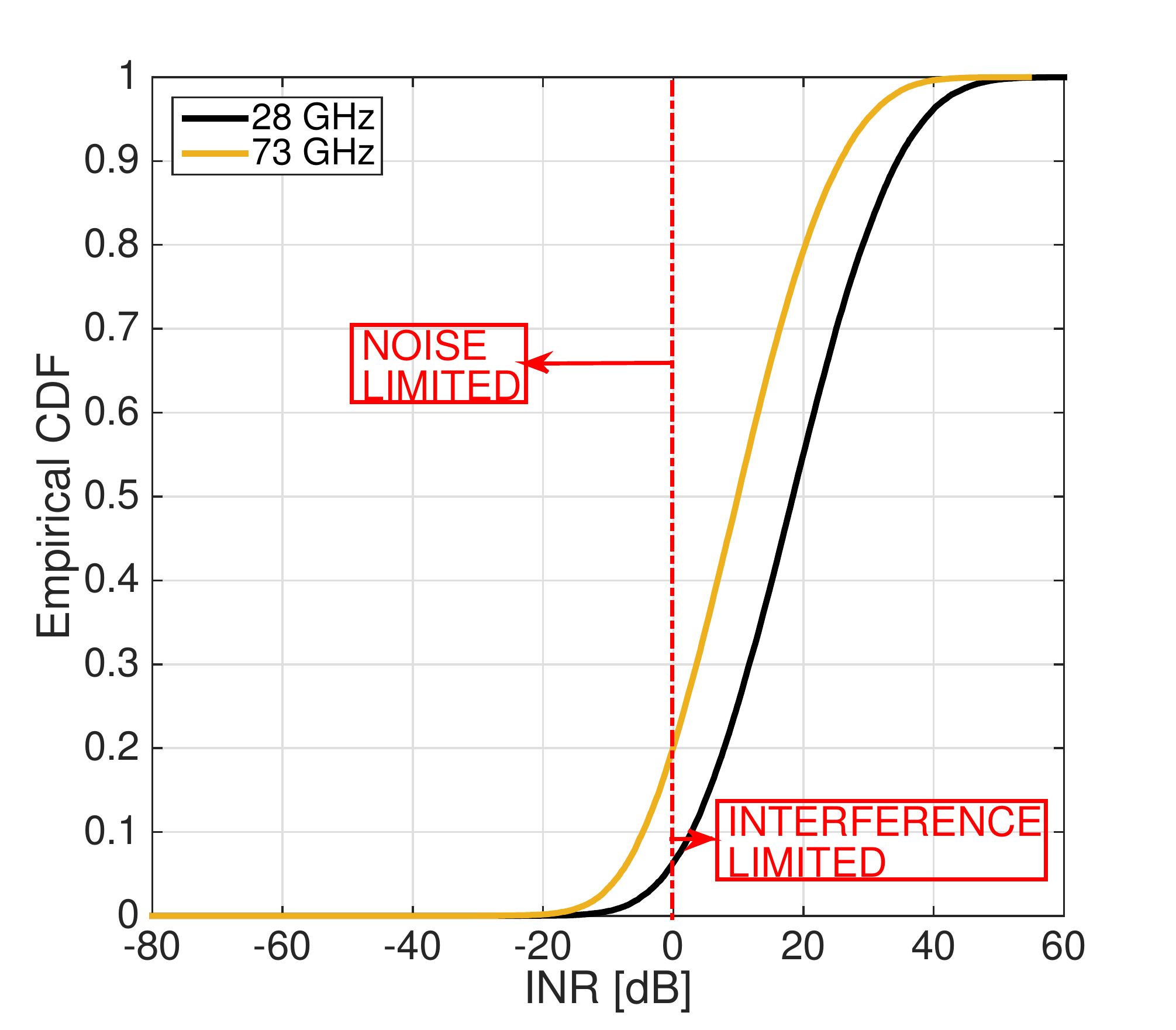}     
\caption{Empirical CDF of the INR for $\lambda_{\text{UE}}$ = 1200~UEs/km$^2$ and $\lambda_{\text{BS}}$ = 120~BSs/km$^2$.}   
\label{1200}
\end{figure}

We consider a 64 elements (8$\times$8) Uniform Planar Array (UPA) antenna at the BS, while at the receiver side we have a UPA with 16 elements (4$\times$4).

For the analysis in this paper, we focus on the INR, which is defined as:
\begin{equation}
\text{INR}_{ij} = \frac{\sum_{k \neq i} \frac{P_{Tx}}{PL_{kj}}G_{kj}}{BW \times N_0},
\label{inr}
\end{equation}
where at the numerator we sum all the interfering links by multiplying their transmit powers $P_{Tx}$, beamforming gains $G$ and respective pathloss values $PL$. The denominator comprises the thermal noise power, which is equal to the power spectral density $N_0$ multiplied by the total bandwidth $BW$. 

We report, for each case, the Empirical Cumulative Distribution Function (ECDF) of the downlink INR defined in~\eqref{inr}, and identify $\text{INR} = 0$ dB as the transitional point that determines the shift from a noise-limited to an interference-limited regime.

In our simulation campaign, we have considered a 7~dB Noise Figure (NF), a 500~MHz total bandwidth, and a transmit power $P_{Tx}=30$~dBm, which are in line with the specifications envisioned for future 5G mmWave mobile networks considering a downlink transmission.

By observing Figures~\ref{300}~--~\ref{1200}, we can derive the following  general insights. 

\subsubsection{Higher BSs density, higher interference} 
A first (obvious) result is that interference increases with the number of base stations, therefore increasingly biasing towards the interference limited regime. We show that we start observing the majority of links ($>$ 80$\%$) operating in the interference limited region for $\lambda_{\text{BS}} =$~120~BSs/km$^2$. 

It is important to note that a large amount of interference is generated from the symmetric lobe.
We reserve as a future work the study of a scenario where an antenna model that blocks symmetric lobes is applied.
We have also verified that changes in the density of the UEs $\lambda_{\text{UE}}$ for fixed $\lambda_{\text{BS}}$ do not affect the behavior of the INR curves. In fact, as long as the number of active BSs (and so the active number of interfering source) remains the same, the behavior of the INR does not change. However, we note that this holds under the assumption of full-buffer UEs.

\begin{figure}[t!]
\centering
\includegraphics[width=0.80\columnwidth]{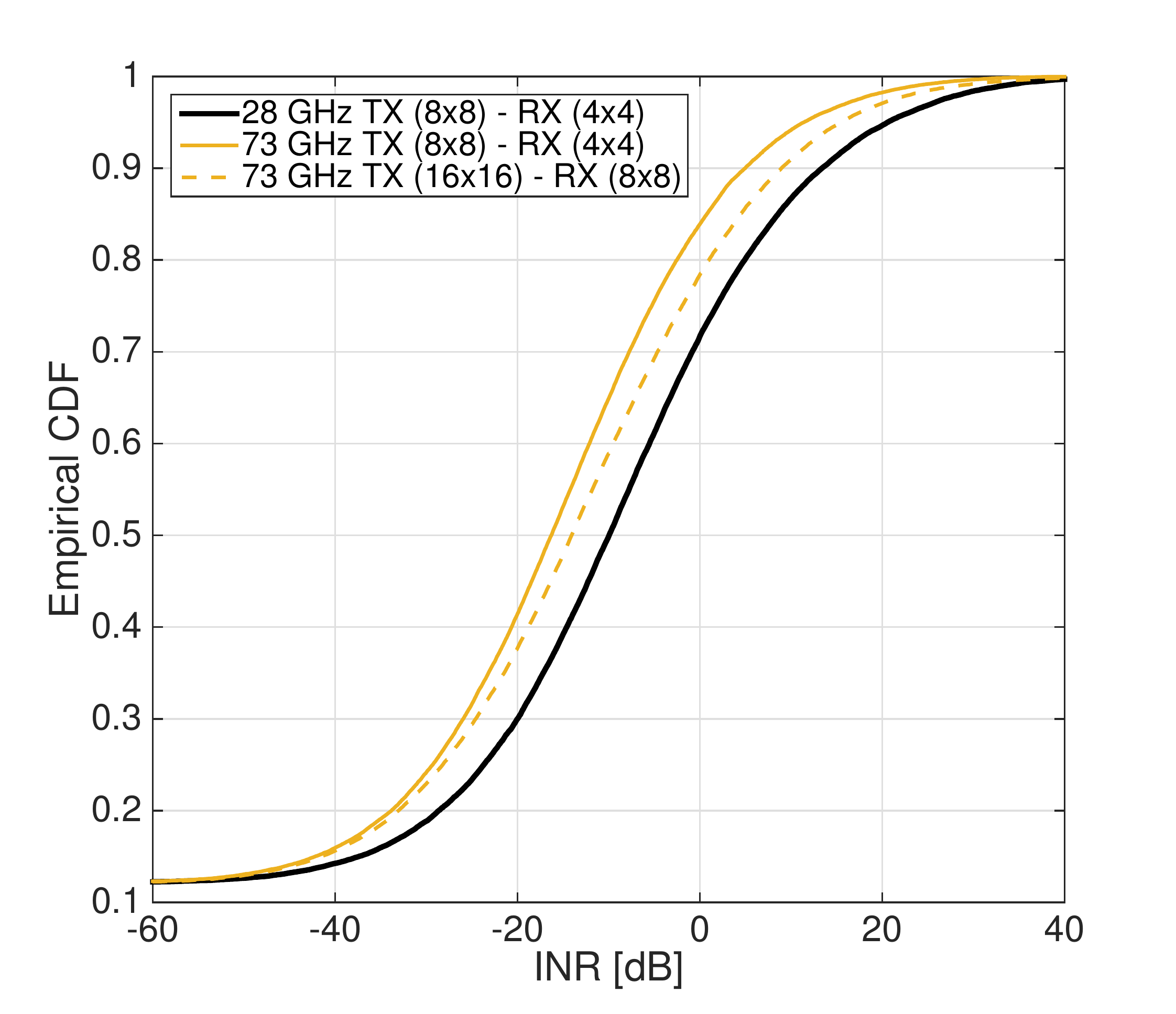}     
\caption{Empirical CDF of the INR for $\lambda_{\text{BS}}$ = 30~BSs/km$^2$ and $\lambda_{\text{UE}}$ = 300~BSs/km$^2$.}
\label{big_antenna_inr}
\end{figure}

\subsubsection{Higher band, lower interference}
 At 73~GHz, due to a higher pathloss, interference is $\sim$ 10~dB less than the interference experienced at lower mmWave bands (28~GHz). Please note that in these first simulations we consider the same number of antennas for both frequencies. Nonetheless, at higher operating bands we could deploy more antennas in the same area, which would result in higher directionality, and therefore even less interference.

\subsubsection{Interferers domains} 
We report in Table~\ref{table_probabilities} the distributions of interfering link states for the 28~GHz curve of Figure \ref{300}.
The table reports the probability for an interferer to be in one of the three states (LoS, NLoS, and outage) for the two ECDF intervals.
We split the curve into two intervals in order to capture the relation between the state of the interferers and the INR value.
\begin{figure}[t!]
\centering
\includegraphics[width=0.80\columnwidth]{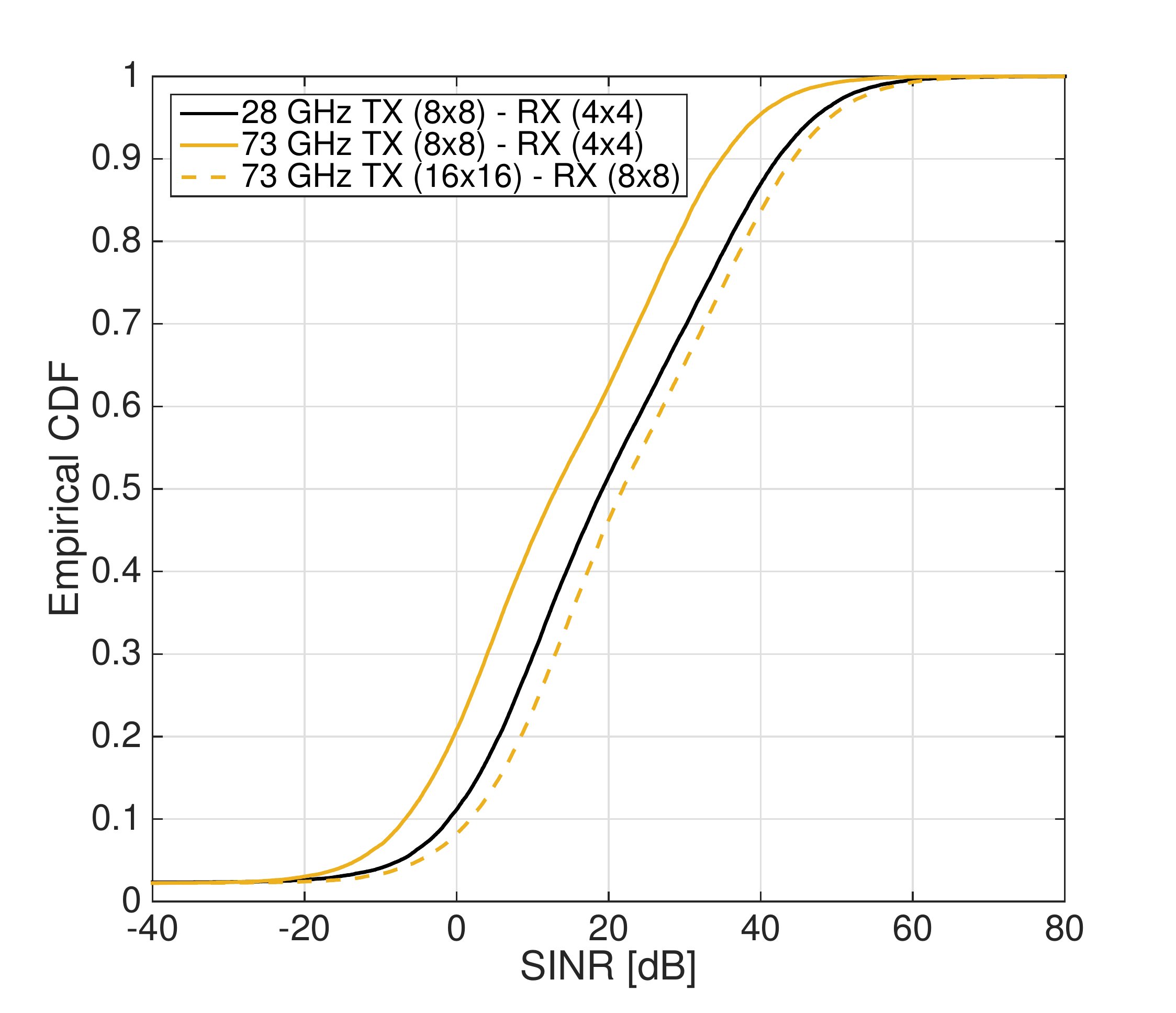}     
\caption{Empirical CDF of the SINR for $\lambda_{\text{BS}}$ = 30~BSs/km$^2$ and $\lambda_{\text{UE}}$ = 300~BSs/km$^2$.}
\label{big_antenna_sinr}
\end{figure}

\begin{table}[h!]
\centering
\begin{tabular}{c|c |c |c }
\bottomrule
 ECDF \emph{interval} & LoS & NLoS & outage \\ [-2pt]\hline
 [0\% - 12\%]            &  0\%               & 0\%                & 100\%                 \\ [-2pt] \hline
[12\% - 100\%]           &  1\%               & 18\%               & 81\%                                \\ [-2pt]\bottomrule
\end{tabular}
\caption{Empirical probabilities to be in LoS, NLoS or outage for the interferers of Figure~\ref{300} (the case in which $\lambda_{\text{UE}}$ = 300~UEs/km$^2$ and $\lambda_{\text{BS}}$ = 30~BSs/km$^2$). We provide the statistics for two ECDF intervals (0~-~12\% and 12~-~100\%), that represent different trends in the plot of Figure~\ref{300}.}
\label{table_probabilities}
\end{table}

It is interesting to note how, for low BS density scenarios, the dominant interfering links are in outage, thus obviously reflecting a noise-limited regime. As observed, at higher BS densities, the states of the interferers become NLoS and LoS, thus increasingly biasing the system towards interference limited regimes.

These results allowed us to identify three main working regimes.
\begin{itemize}
\item When the BS density is smaller than 30~BSs/km$^2$ (average cell radius bigger than 103~m), the mmWave cellular network can be assumed to be \textbf{noise-limited}.
In such a regime, interference coordination will not be necessary. 
\item Conversely, when the density is above 120~BSs/km$^2$, the mmWave cellular network can be assumed to be \textbf{interference-limited}. 
Under this assumption, some sort of interference coordination is needed. 
\item Finally, we captured an intermediate case, where the density is between 30 and 120~BSs/km$^2$; here, we can observe both regimes. 
For this particular state, any user can be either \textbf{noise-limited} or \textbf{interference-limited}. We may need a hybrid coordination scheme in this case. 
\end{itemize}

\subsubsection{More antenna elements at 73~GHz}
So far, we have considered the same antenna array size for both frequencies. Nonetheless, because of the reduced wavelength at 73~GHz, we can deploy a higher number of antenna elements in the same area.  

Figures \ref{big_antenna_inr} and \ref{big_antenna_sinr} report the INR and the SINR ECDFs, respectively. The two cases considered before (solid lines, same number of antennas) are compared to a configuration with an increased number of antennas at 73~GHz, i.e., $16\times16$ at the transmitter and $8\times8$ at the receiver (dashed line). The results show that using more antennas results in a slightly increased interference (about 2--3~dB in most cases, see Figure \ref{big_antenna_inr}). While the average interference due to the main transmit/receive lobes is expected to be roughly the same in the two cases (as the increased gain is compensated by the correspondingly reduced beamwidth), with more antennas the effect of the side lobes is larger, and the interference variance is also larger, thus resulting in an increased INR. On the other hand, the higher directivity of the larger arrays (which in this specific example provides a total gain of 12~dB, i.e., a factor of 4 at both sides) results in an increased SINR value (about 9--10~dB, see Figure \ref{big_antenna_sinr}).
\begin{figure}[t!]
\centering
\includegraphics[width=0.80\columnwidth]{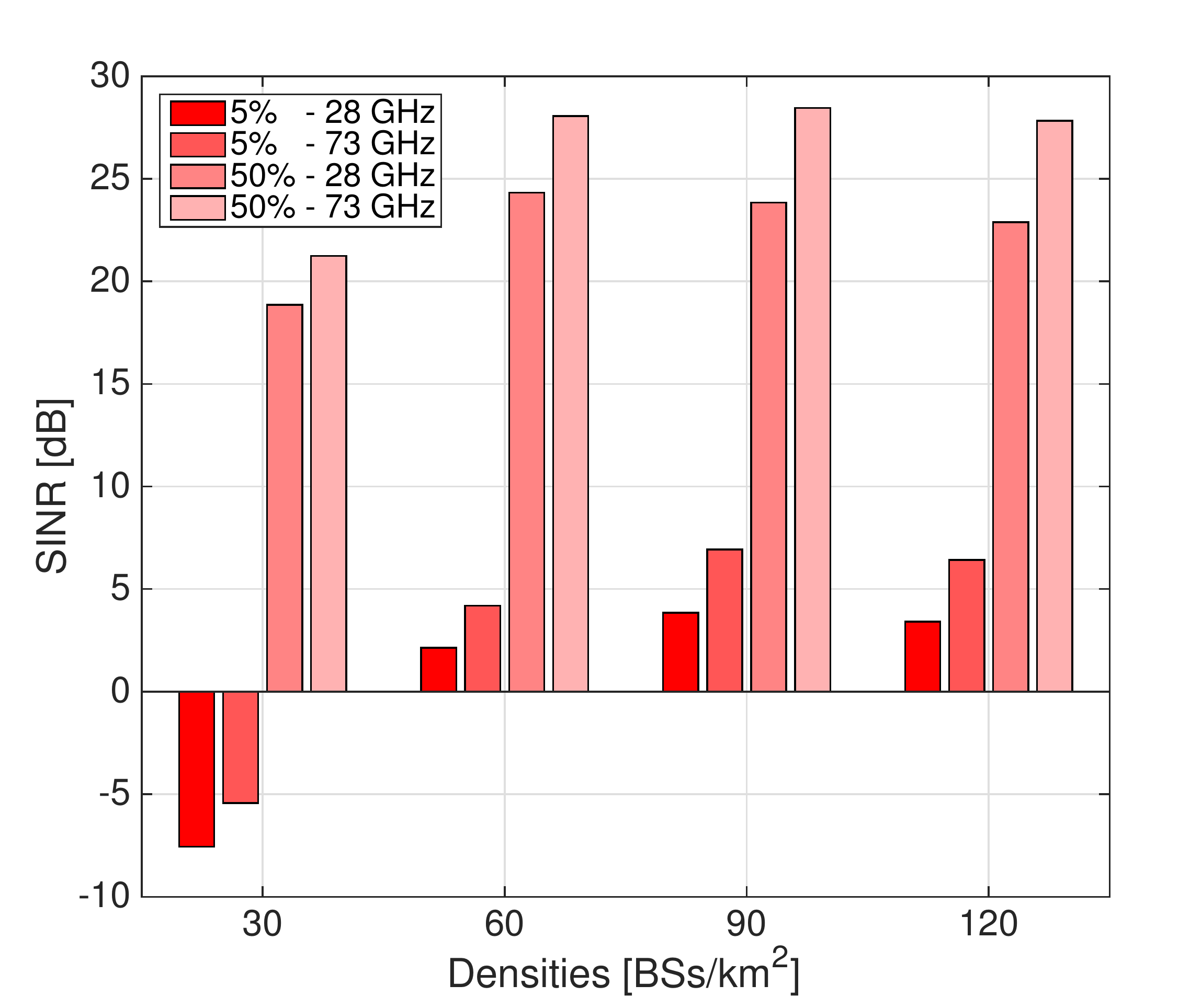}     
\caption{Performance comparison at increasing BS densities for both 28~GHz and 73~GHz.}
\label{comparisons}
\end{figure}

Finally, we report in Figure~\ref{comparisons} the 5th and 50th percentiles of the SINR at 28~GHz and 73~GHz vs. the BS density. An improvement of the SINR when the BS density is increased corresponds to a noise-limited regime, whereas in an interference-limited regime densification leads to a similar increase of both the intended signal and the interference, making the SINR weakly dependent on the BS density (this change of regime occurs between 60 and 90~BSs/km$^2$ for the 5th percentile, and between 30 and 60~BSs/km$^2$ for the 50th percentile). Note that the slight SINR decrease when the BS density is further increased is due to the fact that some interferers move from NLoS to LoS condition, so that their power increases more than that of the intended signal (an effect more visible at 28~GHz).

\section{Conclusion}
\label{conclusion}
This paper leverages the latest measurement-based channel models to accurately assess the interference statistics in a wide range of deployment scenarios. 
The channel models also account for blockage, line-of-sight and non-line-of-sight regimes, as well as local scattering, that significantly affect the level of spatial isolation.

Determining the regime in which the network is operating, and specifically whether it is noise- or interference-limited, is critical in order to properly design MAC and physical-layer procedures.
In this paper, we capture each operating regime as a function of the transmitter density at two different frequencies, i.e., 28~GHz and 73~GHz, and observe their different trends, which depend on directionality and propagation characteristics.

Based on our findings, we believe that a flexible, user-centric, interference-aware MAC protocol may represent the right solution to better leverage the mmWave potential, which motivates us to further investigate these possibilities as part of our future work.

\bibliographystyle{IEEEtran}
\bibliography{biblio}

\end{document}